\documentclass[a4paper,11pt]{article}
\usepackage{jinstpub} 
\usepackage{lineno}
\RequirePackage{graphicx}
\RequirePackage[version=4]{mhchem}
\RequirePackage{siunitx}

\RequirePackage{amssymb}
\RequirePackage{mathptmx}     
\RequirePackage{hyperref}
\usepackage{wasysym}
\usepackage[symbol]{footmisc}
\usepackage{anyfontsize}
\usepackage{booktabs}
\usepackage{caption} 
\usepackage{amsmath}

\title{ASTAROTH: A Novel Detector for Dark Matter Direct Detection Using Cryogenic SiPMs}

\author[a,*]{E.~Martinenghi\note{Corresponding author.}}
\author[a,b]{, V.~Toso}
\author[a,b]{, F.B.~Armani}
\author[a,c]{, A.~Castoldi}
\author[d]{, G.~Di Carlo}
\author[a]{, L.~Frontini}
\author[a,b,e]{, N.~Gallice}
\author[a,c]{, C.~Guazzoni}
\author[a,b]{, V.~Liberali}
\author[b]{, L.~Rutigliani}
\author[a,b]{, A.~Stabile}
\author[d]{, K.~Szczepaniec}
\author[a,b]{, V.~Trabattoni}
\author[a]{, A.~Zani}
\author[a,b]{and D.~D'Angelo}

\affiliation[a]{INFN - Sezione di Milano, via Celoria 16, 20133 Milano, Italy}
\affiliation[b]{Dipartimento di Fisica, Universit\`a degli Studi di Milano, via Celoria 16, 20133 Milano, Italy}
\affiliation[c]{Dipartimento di Elettronica, Informazione e Bioingegneria (DEIB), Politecnico di Milano, piazza Leonardo da Vinci 32, 20133 Milano, Italy}
\affiliation[d]{INFN - Laboratori Nazionali del Gran Sasso (LNGS), via G. Acitelli 22, 67100 Assergi, Italy}
\affiliation[e]{Brookhaven National Laboratory, PO 5000, Upton, NY 11973, USA}

\emailAdd{edoardo.martinenghi@mi.infn.it}

\abstract{The DAMA experiment’s long-standing claim of dark matter detection remains a key open issue in astroparticle physics. Independent verification requires NaI(Tl)-based detectors with enhanced low-energy sensitivity. Current detectors rely on photomultiplier tubes (PMTs) which features limited detection efficiency, intrinsic radioactivity, and high noise at few-keV energies.
ASTAROTH is an R\&D project that developed a proof of concept NaI(Tl) detector where silicon photomultipliers (SiPMs) have been used instead of PMTs, offering higher photon detection efficiency, negligible radioactivity, and, most of all, a reduction of two orders of magnitude in the dark noise. The setup includes a custom cryostat operating at approximately 80 K.
We report the first characterization of an approximately 360~g NaI(Tl) crystal coupled to a 5×5 cm² SiPM matrix, yielding 4.5 photoelectrons/keV after crosstalk correction. This promising result demonstrates the feasibility of SiPM-based readout for NaI(Tl) and paves the way for future large-scale dark matter experiments.}

\begin{document}
\maketitle
\flushbottom

\section{Introduction}
\label{sec:intro}
The nature of dark matter remains one of the open questions in modern physics. A number of astronomical observations \cite{Sofue2001,Clowe2006,Freese2009} can be explained by the introduction of dark matter \cite{Bertone2010}, and several candidates have been proposed. A major scientific effort focuses on the detection of Weakly Interacting Massive Particles (WIMPs)~\cite{Goodman1985}, with numerous experiments aiming to observe their recoil energy from interactions with target nuclei. The expected recoil energy is of the order of a few-keV and occurs at very low event rates~\cite{Lewin1996}. Therefore, the design of detectors with extremely low background and low detection thresholds is essential, representing one of the main challenges in direct detection experiments.

To date, the only positive result indicating a possible dark matter interaction has been reported by the DAMA experiment \cite{Bernabei2013, Bernabei2021}, which observed an annual modulation in the event rate of an array of NaI(Tl) scintillating crystals located underground at the Laboratori Nazionali del Gran Sasso (LNGS) in Italy. Despite being of extreme impact, this result has been excluded by several more sensitive experiments (e.g., XENON~\cite{Xenonnt}, PandaX~\cite{PandaX2025}, LZ~\cite{LZ2024}, DarkSide~\cite{Agnes2018_2}). However, since these experiments employ different target materials, a possible dependency of WIMPs interaction with target nuclei has been hypothesized \cite{Catena2016}. Consequently, a number of NaI(Tl)-based experiments have been carried out (e.g., ANAIS \cite{anais2025}, COSINE \cite{cosine2025}) or are under preparation (e.g., SABRE \cite{Calaprice2022}, PICOLON \cite{Fushimi2022}) to provide a definitive answer.

These experiments share a common design in which the crystals are coupled to photomultiplier tubes (PMTs). Although PMTs are a well-established technology, they present intrinsic limitations for dark matter detection. PMTs exhibit reduced photon detection efficiency (PDE) at the NaI(Tl) emission wavelength (e.g., the Hamamatsu R11065-20 has about 35\% PDE at 420~nm) and, most importantly, they have high intrinsic noise, which increases the experimental background and limits the achievable signal-to-noise ratio (SNR) at low energies \cite{Mariani21} (i.e., below 1~keV). An alternative solution is the use of silicon photomultipliers (SiPMs) \cite{Dinu2016} instead of PMTs, which offer a higher PDE (about 65\% at 420~nm ~\cite{Merzi_2023}) and a reduction of dark noise by two orders of magnitude when operated at cryogenic temperatures (down to about 80~K) ~\cite{Biroth2015}.

With the ASTAROTH project, we aim to replace PMTs with cryogenically operated SiPMs, demonstrating the suitability of this technology for dark matter direct detection experiments. We present here the first ASTAROTH prototype, located at LASA (Laboratorio di Acceleratori e Superconduttività Applicata) in Segrate, Italy. In this paper, we describe the adopted technological solutions and report the first experimental results obtained. ASTAROTH is intended to serve as a crucial building block toward the design of larger-mass experiments capable of exploring the sub-keV energy region, where new information may yet be uncovered.

\section{Experimental Setup}

\subsection{Cryostat}
\begin{figure}[t]
\centering
\includegraphics[width=.6\textwidth]{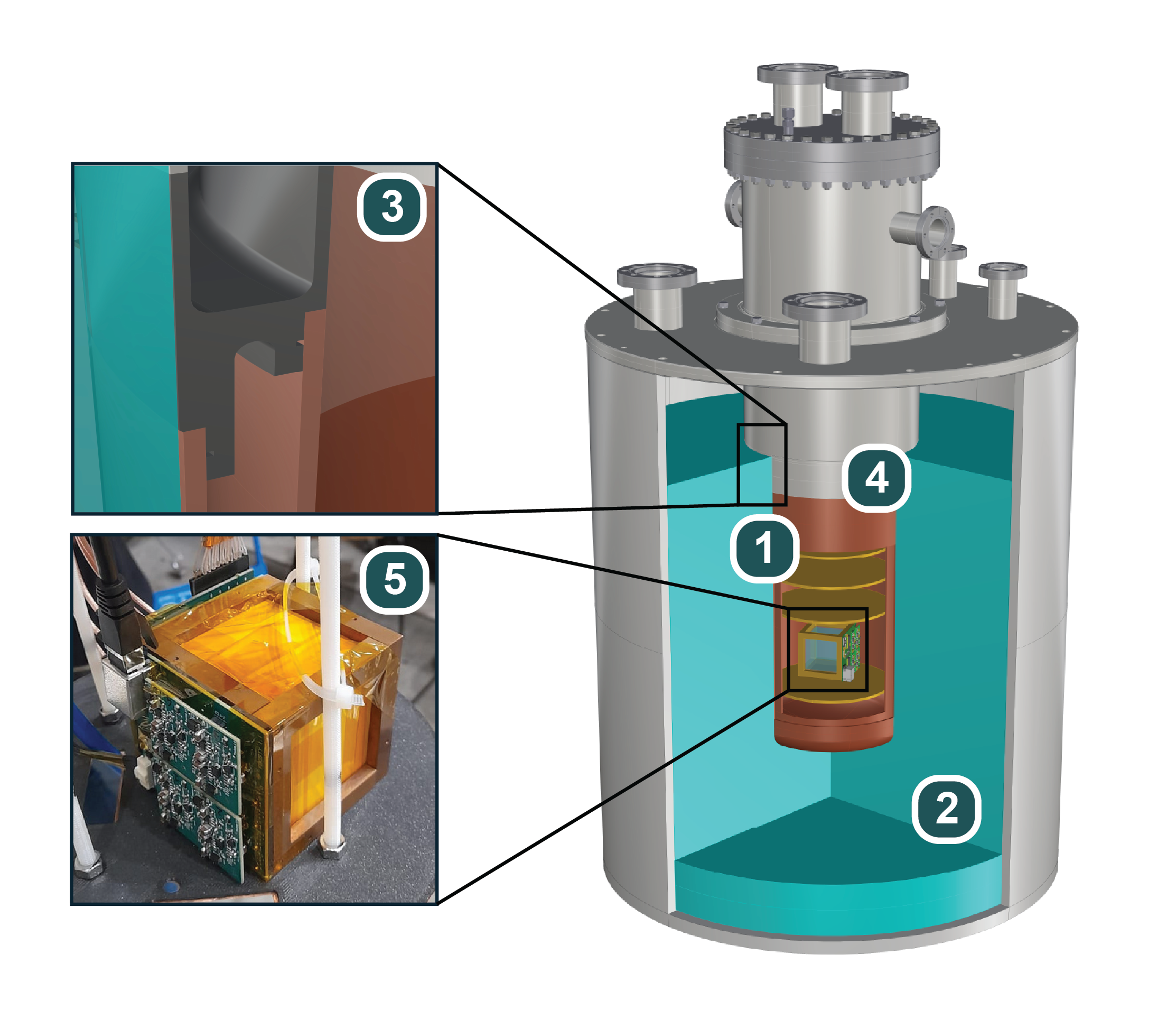}
\qquad
\caption{Design of the cryostat. The cryogenic chamber (1) is submerged into a cold fluid (2) exchanging heat through the thermal bridges (3). The detector (5) is hosted inside the chamber, which is connected to the chimney (4).
\label{fig:cryostat}}
\end{figure}
To operate the detector at cryogenic temperature, an innovative cryostat has been designed. The cryostat operates at a tunable temperature between \qtyrange[range-units=single]{80}{150}{\kelvin}. The upper limit corresponds to the temperature where the SiPM dark count rate is equal to the typical PMT noise ($\approx$~0.1~cps/mm$^2$) \cite{SABRESouthPMTS2025}, while the lower limit is set by the cryogenic fluid temperature. The cryostat was developed through extensive simulations and cryogenic stress tests performed by the Mechanical Design Services of INFN Milan and INFN LNGS, in collaboration with the authors~\cite{Zani_2021}. 
At \qty{80}{\kelvin}, the FBK SiPMs already reach DCR values of about \qty[print-unity-mantissa=false]{e-3}~cps/mm$^2$, with negligible improvement expected at lower temperatures~\cite{Acerbi2017}.

Figure \ref{fig:cryostat} shows a design of the cryostat, where cooling is provided by a cold fluid (2) circulating around a cryogenic chamber (1) that houses the detector (5). The chamber consists of a double-walled Oxygen-Free High-Conductivity (OFHC) copper vessel with an inner diameter of 214~mm, joined to a stainless-steel (SS) chimney (4). OFHC copper ensures high thermal conductivity and low intrinsic radioactivity. The copper section features two vacuum-insulated 3~mm-thick walls brazed to a stainless-steel bridge (3). This bridge provides the main heat transfer path, as radiative exchange between the walls is negligible. The inside of the chamber is filled with helium gas to act as heat transfer medium between the chamber walls and the detector. A set of thin copper disks along the chamber height enhances gas stratification, suppressing convective flows. In the current setup, liquid nitrogen (i.e. \qty{77}{\kelvin}) was used as cryogenic fluid.

\subsection{NaI(Tl) Crystal}

\begin{figure}[t]
\centering
\includegraphics[width=.5\textwidth]{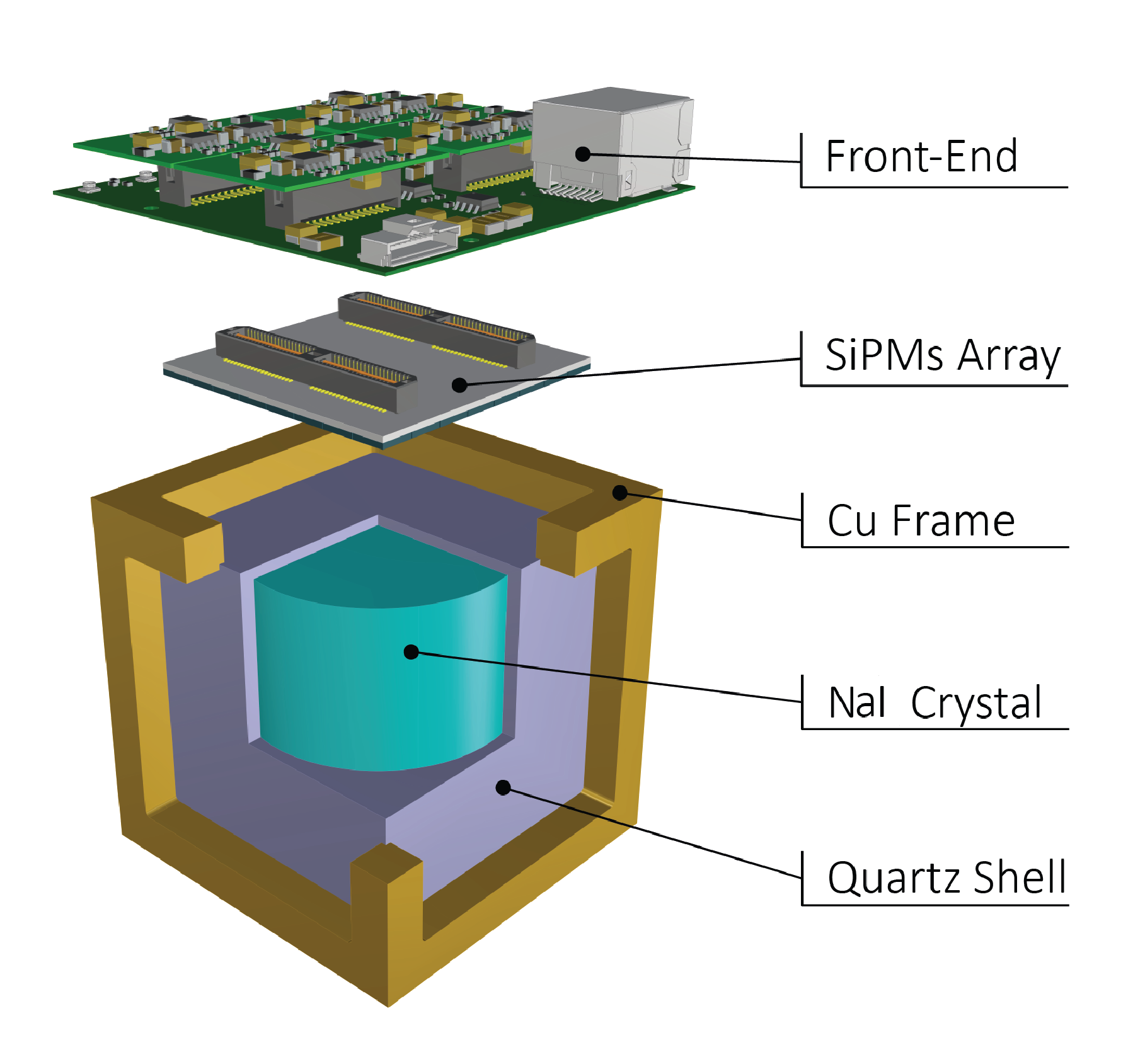}
\qquad
\caption{Design of the detector. A SiPM array is coupled to the crystal by means of a copper frame, the electronic front-end reads out the signal from the SiPM.
\label{fig:detector}}
\end{figure}

The detector employed in this work consists of a cylindrical NaI(Tl) crystal (750~ppm thallium doping, Hilger Crystals, UK) coupled to a SiPM matrix on one face, as illustrated in Fig.~\ref{fig:detector}. The crystal has both diameter and height of 50~mm. Its emission spectrum peaks at \SI{420}{\nano\metre}, matching the sensitivity range of NUV-class SiPMs. The typical light yield at room temperature is about 40 photons/keV with a main de-excitation time of \qty{250}{\nano\second}~\cite{Knoll2010}, however, at cryogenic temperature, significantly higher response times (i.e. up to \qty{1.5}{\micro\second}) have been observed~\cite{Sibczynski2011,Gallice23}.

To prevent hygroscopic degradation, the crystal is sealed in an airtight fused-silica shell transparent at \SI{420}{\nano\metre}, filled with dry neon gas at 1~bar. A uniform 1~mm gap between crystal and enclosure accommodates thermal expansion differences through cryogenic cycles. The cubic shell (\qtyproduct{58 x 58 x 58}{\milli\metre}) is surrounded by an OFHC copper frame ensuring mechanical support and coupling to the SiPM and readout electronics.

Although this configuration, suggested by the industrial partner for ease of production, offers mechanical robustness and low leakage risk, the total internal reflection at the crystal–neon interface limits the light emission. A Geant4-based Monte Carlo simulation estimated that only about \qty{35}{\percent} of the emitted photons reach the photosensors~\cite{Galli2022}. Therefore, an improved coupling scheme based on epoxy resin coating is under development to mitigate this effect.

\subsection{SiPM and Electronics}
NUV-HD-cryo low-field SiPMs developed by Fondazione Bruno Kessler (FBK) and optimized for cryogenic operation were employed in the present detector. Each SiPM measures \qtyproduct{6 x 6}{\milli\metre} with a \SI{30}{\micro\metre} cell pitch. When operated in liquid nitrogen (77~K) at \SI{7}{\volt} excess-bias, they exhibit a PDE of \qty{55}{\percent} at \SI{420}{\nano\metre}, a dark count rate below 10$^{-2}$~cps/mm$^{2}$, and afterpulsing probability under 20\%~\cite{Acerbi2017}.

The matrix, assembled by FBK following a jointly developed design, includes 64 SiPMs arranged in an \numproduct{8 x 8} configuration on a single printed circuit board (PCB), providing a total active area of \qtyproduct{48 x 48}{\milli\metre}. Electrical connections are made via conductive glue (cathodes) and wire bonding (anodes). The breakdown voltage spread is below 0.5\%, ensuring uniform performance. A thin (i.e. $\sim$ \SI{100}{\micro\metre}) epoxy resin layer protects the SiPMs during coupling with the crystal. Two high-pitch 80-pin connectors on the back of the PCB link the array to the front-end electronics.

The custom front-end electronics comprise a \textit{cold section}, located inside the cryogenic chamber and directly coupled to the SiPMs, and a \textit{warm section} outside the cryostat. The design employs SiGe integrated circuits tested for cryogenic operation, together with metal-film resistors and C0G capacitors, known for thermal stability down to 4~K~\cite{Pan2005}. Circuit optimization was supported by SiPM modeling and dedicated simulations~\cite{Acerbi2019,Villa2015}.

The cold front-end implements a three-stage amplification chain for the 64 SiPM channels. The first stage uses LMH6626-based transimpedance amplifiers (TIA) summing the output of four SiPMs each. A second LMH6624-based stage combines these into four quadrant signals (Q1–Q4), each corresponding to 16 SiPMs. A final stage, using THS4541 fully differential op-amps, converts the analog signals to differential pairs transmitted via shielded CAT7 cables to the \textit{warm section}. There, the signals are reconverted to single-ended for acquisition, which employs an 8-channel, 14-bit, 500~MS/s digitizer (CAEN V1730) \cite{caenv1730}.

\section{Experimental Results}
\label{sec:exp-results}
\subsection{System Calibration}

 \begin{figure}[t]
\centering
\includegraphics[width=1\textwidth]{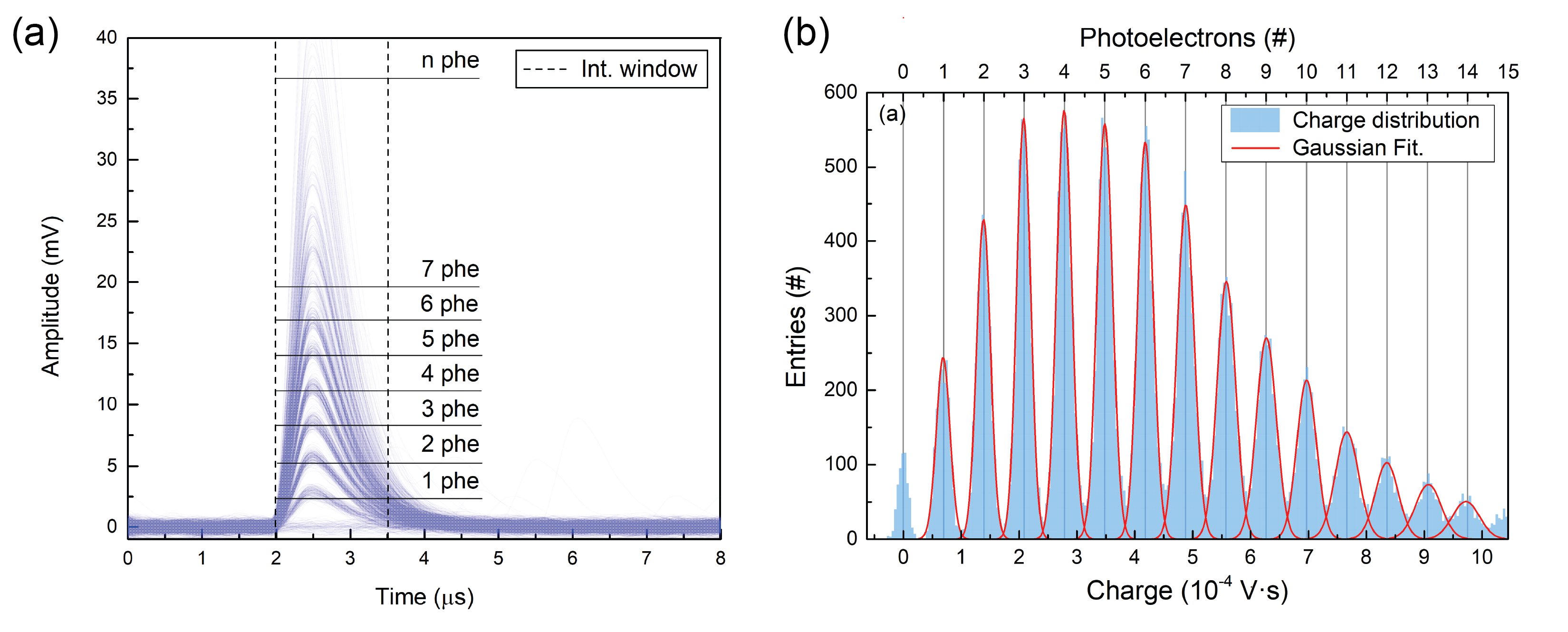}
\qquad
\caption{(a) Baseline subtracted waveforms from the SiPM output, the amplitude of the pulse linearly increases with the number of detected photons. Integration window for the charge computation is indicated between dashed lines. (b) Histogram of charge corresponding to detected photon with superimposed Gaussian fitting for each peak.}
\label{fig:calibration}
\end{figure}

\begin{figure}[t]
\centering
\includegraphics[width=1\textwidth]{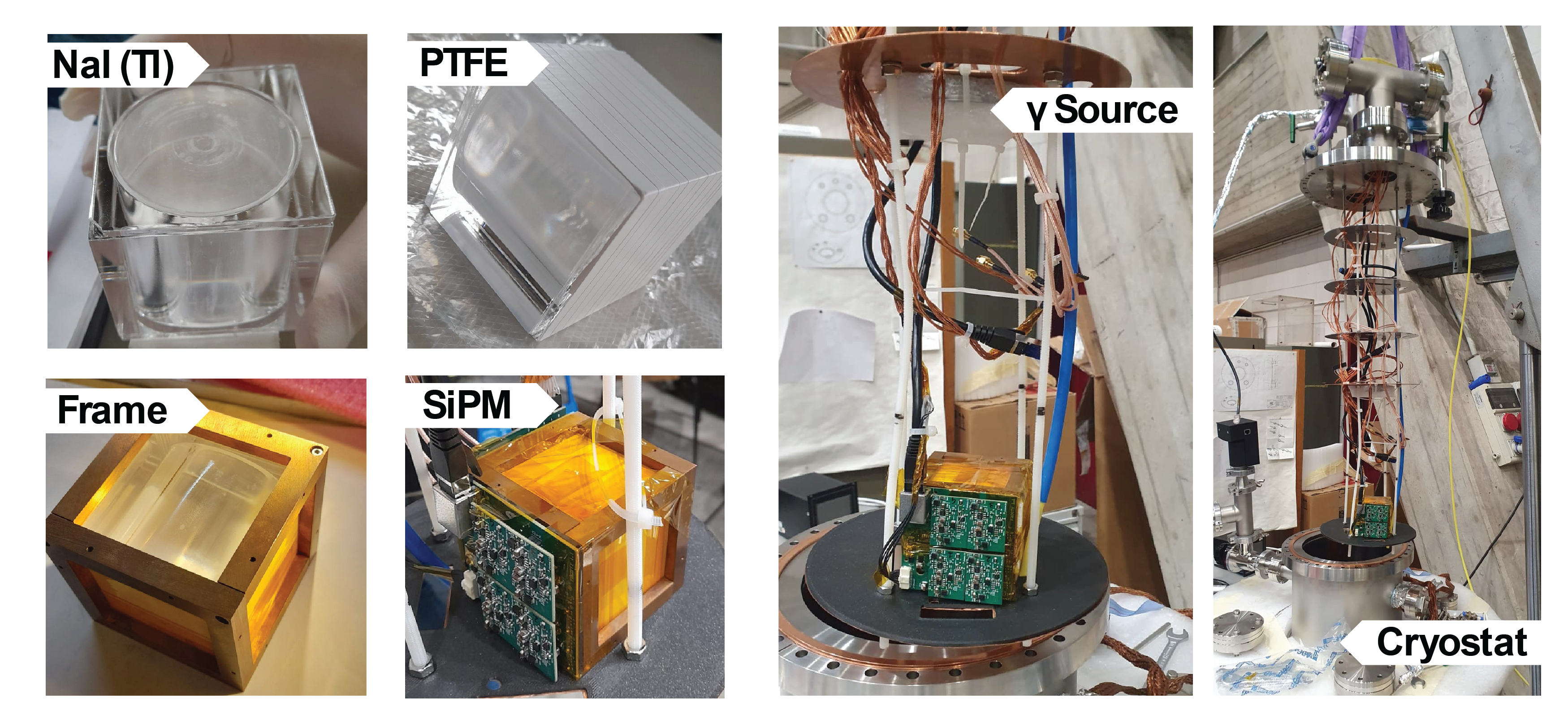}
\qquad
\caption{Pictures of the experimental setup, the NaI(Tl) crystal is covered into PTFE tape for increasing the diffuse reflectivity onto the non-instrumented faces and then, thanks to a copper frame, coupled to the SiPM. After positioning a $^{241}$Am source at 25 cm distance, the detector is installed into the cryostat and cooled down to $\approx$ 80~K.
\label{fig:picture}}
\end{figure}

The SiPM matrix and front-end electronics were characterized and calibrated prior to coupling with the crystal. To reproduce cryogenic conditions, the SiPM and the \textit{cold section} of the electronics were operated in evaporating \ce{LN2} at \SI{77}{\kelvin}, inside a light-tight Faraday cage.

A picosecond-pulsed 405~nm laser (Hamamatsu M10306-30 + C10196 controller~\cite{hamamatsu}) was used for calibration in the single-photon regime. The laser repetition rate (\textless100~Hz) avoided noise typical of cryogenic SiPMs such as afterpulsing ~\cite{Acerbi2017}. The SiPM output was recorded for \num{40000} laser repetitions and baseline subtraction was applied to each waveform.

As reported in Figure~\ref{fig:calibration}~(a), single-photoelectron pulses shows amplitudes of about \qty{2.5}{\milli\volt}, with linear scaling for multiphoton events. The signal, shaped by parasitic capacitance and front-end bandwidth, features rise and decay times of \qty{0.2}{\micro\second} and \qty{0.5}{\micro\second}, respectively. Charge calibration was obtained by integrating over a \SI{1.5}{\micro\second} window, capturing over 95\% of the pulse area.

The charge distribution histogram (Figure~\ref{fig:calibration}~(b)) exhibited well-separated photon peaks, fitted with Gaussian functions. The linearity of the response was confirmed with $R^2 = 0.99999$ allowing to precise derivate the calibration factor from integrated charge into number of detected photoelectrons. A crosstalk probability of $p = 0.37$ was derived using the Vinogradov method~\cite{vinogradov2009,Vinogradov2012} and used for photoelectron yield correction as described in the following section.

\subsection{Scintillation Light Detection}

\begin{figure}[t]
\centering
\includegraphics[width=0.6\textwidth]{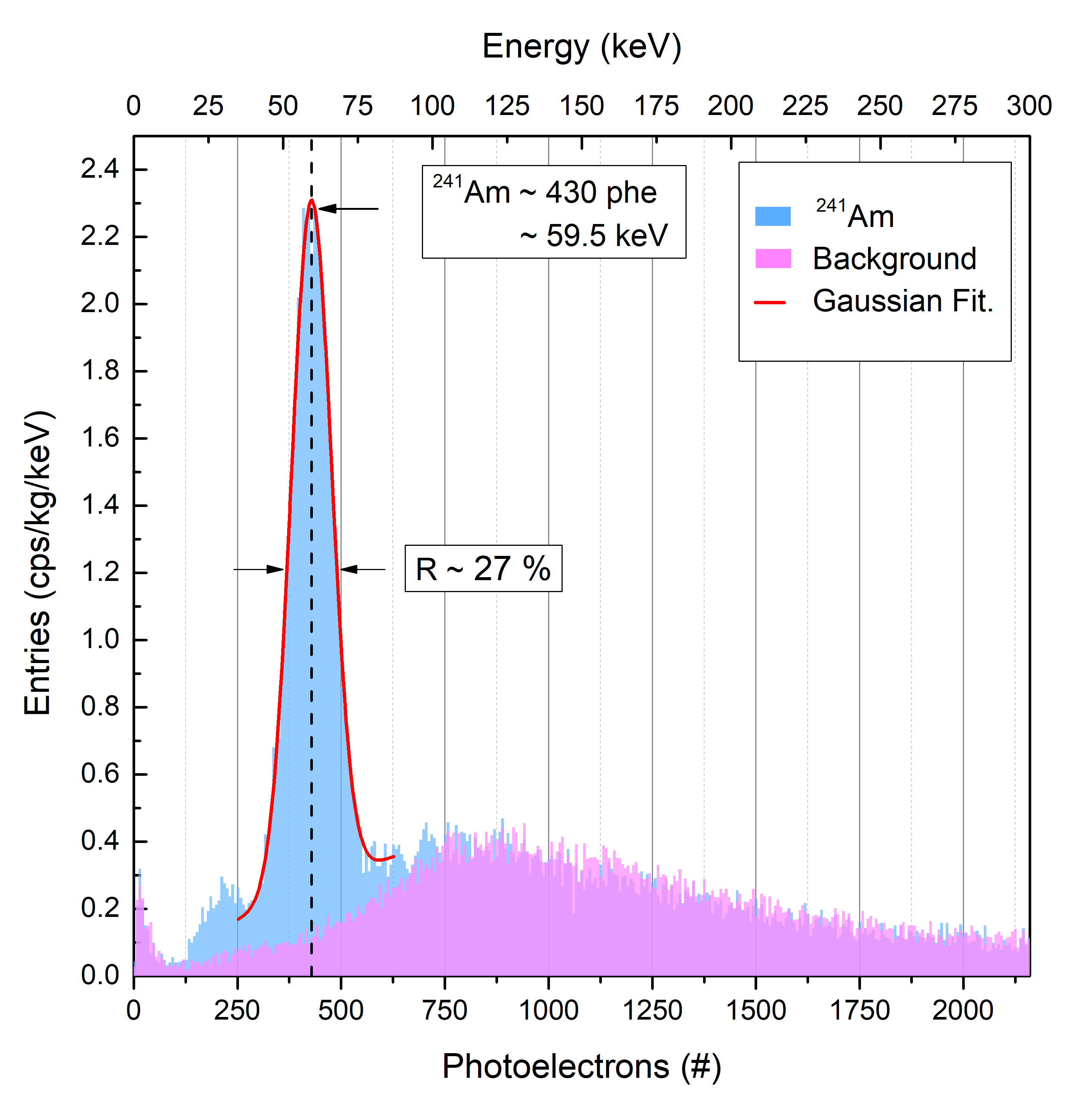}
\qquad
\caption{Energy spectra comparing two parallel run performed with the \textsuperscript{241}Am source (blue) and without (magenta). The peak stands at 59.5~keV due to gammas from the source. A Gaussian fitting returns the peak position and a gross photoelectron yield of 7.2~phe/keV (4.5~phe/keV after crosstalk correction). The upper x-axis is computed consequently.\label{fig:source}}
\end{figure}

\label{sec:Detector}

After calibration, the SiPM matrix and the \textit{cold section} of the electronics were coupled to the crystal as shown in Figure~\ref{fig:picture}. The NaI(Tl) crystal was wrapped in PTFE tape to enhance light collection through diffuse reflection. The crystal was then mounted onto the copper frame and coupled to the SiPM on one face. Finally the detector was coupled to a \SI{47}{\kilo\becquerel} \ce{^{241}Am} source, placed at about 25~cm distance, and mounted inside the cryostat. The cryostat external tank was then filled with liquid nitrogen, cooling the internal chamber to about \SI{80}{\kelvin}. 

For data acquisition, each quadrant’s threshold was set to \qty{1.5}{\milli\volt}, well above the RMS noise (\qty{0.5}{\milli\volt}), and a 4-way coincidence within \qty{48}{\nano\second} was implemented on the four quadrants of the SiPM, using the CAEN V1730 digitizer to isolate scintillation events from noise. A \qty{12}{\micro\second} acquisition window was used.

For each run, 40,000 waveforms were integrated over a \qty{5}{\micro\second} window to account for the NaI(Tl) scintillation decay time constant, which at cryogenic temperature is expected to be in the order of \qty{1.5}{\micro\second}~\cite{Sailer2012,Sibczynski2011}. Charges were converted to photoelectrons using the calibration already performed, and summed over the four quadrants.

Figure~\ref{fig:source} shows the spectrum acquired when the detector is coupled to the radiactive source (blue) and, as comparison, a measurement performed when only the radioactive background of the setup is taken into account (magenta). The 59.5~keV $\gamma$ peak from $^{241}$Am corresponds to 430~photoelectrons, yielding 7.2~phe/keV. After correcting for crosstalk ($p=0.37$), the net yield is 4.5~phe/keV, with an energy resolution of (26.6~$\pm$~0.4)\% FWHM at 59.5~keV. The trigger condition of four photoelectrons corresponds to $\sim$0.5~keV. The structure observed between 20–40~keV is attributed to decay sub-products of \ce{^{241}Am} (e.g., \ce{^{237}Np}) which the system isn't able to discriminate due to lack of energy resolution.

\section{Summary and Outlook}
ASTAROTH is an R\&D program, aiming to demonstrate the possibility of replacing PMTs with large-area cryogenic SiPM matrices to enhance the signal-to-noise ratio at keV-scale energies in dark matter direct detection experiments. The target design employs cubic NaI(Tl) crystals (5~cm side) read out on all faces by SiPMs operating at 80–150~K in a custom cryostat, where dark noise is strongly suppressed.

This work reports measurements performed at LASA with a prototype detector comprising a cylindrical NaI(Tl) crystal (5~cm $\times$ 5~cm) coupled to a single SiPM matrix. Using a \ce{^{241}Am} source, a net photoelectron yield of 4.5 phe/keV was obtained after crosstalk correction.

Next steps include: (1) implementation of cubic epoxy-encased crystals to prevent signal loss of the current setup; (2) optical coupling of SiPMs to multiple faces to increase the photoelectron yield; (3) adoption of liquid argon as both coolant and active veto medium~\cite{Heindl_2010,Doke_2002,cosine2021veto,SABREPoP} to reject background radioactivity; and (5) operation in an underground laboratory to avoid disruptive interaction with cosmic muons. ASTAROTH represents a significant step toward scalable, low energy threshold and low-radioactivity detectors for the next-generation of NaI based dark matter direct detection experiments.

\acknowledgments

This work has been supported by the Fifth Scientific Commission (CSN5) of the Italian National Institute for Nuclear Physics (INFN), through the ASTAROTH grant.




\bibliographystyle{spphys}       
\bibliography{biblio.bib}   





\end{document}